\documentclass[5p]{elsarticle}

\usepackage{graphicx}
\usepackage{epsf} 
\usepackage{amsfonts}
\usepackage{amsmath}
\usepackage{amssymb}
\usepackage{bbold}
\usepackage{bm}
\usepackage{bbm}
\usepackage{slashed}
\usepackage{dcolumn}
\usepackage{color}
\usepackage[caption=false]{subfig} 

\begin{document} 

\title{Covariant Spectator Theory of heavy-light and heavy mesons and the predictive power of covariant interaction kernels}

\author[cftp]{Sofia Leit\~ao}
\ead{sofia.leitao@tecnico.ulisboa.pt}

\author[ue,cftp]{Alfred Stadler\corref{cor1}}
\ead{stadler@uevora.pt}

\author[dfis,cftp]{M. T. Pe\~na}
\ead{teresa.pena@tecnico.ulisboa.pt}

\author[cftp]{Elmar P. Biernat}
\ead{elmar.biernat@tecnico.ulisboa.pt}

\address[cftp]{CFTP, Instituto Superior T\'ecnico, Universidade de Lisboa, Av.\ Rovisco Pais 1, 
1049-001 Lisboa, Portugal}
\address[ue]{Departamento de F\'isica, Universidade de \'Evora, 7000-671 \'Evora, Portugal}
\address[dfis]{Departamento de F\'isica, Instituto Superior T\'ecnico, Universidade de  Lisboa, Av.\ Rovisco Pais 1, 
1049-001 Lisboa, Portugal}

\cortext[cor1]{Corresponding author}


\begin{abstract}
The Covariant Spectator Theory (CST) is used to calculate the mass spectrum and vertex functions of heavy-light and heavy mesons in Minkowski space. The covariant kernel contains Lorentz scalar, pseudoscalar, and vector contributions. The numerical calculations are performed in momentum space, where special care is taken to treat the strong singularities present in the confining kernel. The observed meson spectrum is very well reproduced after fitting a small number of model parameters. Remarkably, a fit to a few pseudoscalar meson states only, which are insensitive to spin-orbit and tensor forces and do not allow to separate the spin-spin from the central interaction, leads to essentially the same model parameters as a more general fit. This demonstrates that the covariance of the chosen interaction kernel is responsible for the very accurate  prediction of the spin-dependent quark-antiquark interactions.
\end{abstract}

\begin{keyword}
Covariant quark model \sep meson spectrum \sep covariant spectator theory
\PACS 14.40.-n \sep 12.39.Ki \sep 11.10.St \sep 03.65.Pm
\end{keyword}

\maketitle
The rigorous calculation of hadronic bound states in QCD is still an open problem. It is hoped that, eventually, lattice QCD will explain all observed hadrons in terms of quark and gluon degrees of freedom. Nevertheless, models have played---and will continue to do so---an important role in aiding the extrapolations of lattice QCD results to physical quark masses, but also in the interpretation of the experimental data and in the analysis of different dynamical mechanisms.

The physics of mesons, in particular,  is a very active area of research, especially due to the ample amount of new experimental data measured at facilities such as the LHC, BaBaR, Belle, CLEO, and more exciting results can also be expected from Jefferson Lab (GlueX) and FAIR (PANDA) in the near future. Some of the recently discovered states (e.g., the X(3872)  in charmonium \cite{Choi:2003zl}) have surprising properties that seem incompatible with an interpretation as $q\bar{q}$ states, sparking particular interest from theorists.

The purpose of this work is twofold: First, we present results of relativistic calculations of $q\bar{q}$ bound states for systems with at least one heavy ($b$ or $c$) quark using the manifestly covariant framework of the Covariant Spectator Theory (CST) \cite{Gro69,Gross:1982,Sta11}. Second, we show that our covariant kernel correctly predicts the spin-dependent interactions when it is fitted to data that do not contain any independent information about them. More precisely, when the kernel is fitted exclusively to pseudoscalar meson states, which are $S$-waves and thus insensitive to spin-orbit and tensor forces (and which do not allow to isolate the spin-spin interaction because here it always acts on singlets), the vector, scalar and axial-vector states which \emph{do} depend on them are correctly described. We believe that this is an important test, performed here for the first time, which confirms the predictive power of covariant kernels.

Most quark models are variations of the nonrelativistic Cornell potential \cite{Eichten:1975bs} which consists of a short-range color-Coulomb and a linear confining potential and was surprisingly successful in describing heavy quarkonia. Because light quarks require a relativistic description, in order to be applicable to all $q\bar{q}$ states these Cornell-type potentials were ``relativized'' \cite{Godfrey:1985} by including a number of relativistic corrections. For a more rigorous treatment of relativity, a number of relativistic equations related to the Bethe-Salpeter equation (BSE) were applied to calculate the meson spectrum \cite{TIEMEIJER199238,Spence:1993bh}, and, more recently, also covariant two-body Dirac equations in the framework of constraint Hamiltonian dynamics \cite{Crater:2010pi,Giachetti:2013uo} gave very good results. The Lorentz structure of the confining interaction in these approaches is not quite settled, although in most cases a scalar structure dominates.

The influential Dyson-Schwinger-Bethe-Salpeter (DS-BS) approach \cite{Burden:1997vl,Maris:1999dk,Holl:2005wq,Hilger:2014nma} is also covariant, but implements confinement not through a confining interaction but through the requirement that there be no real mass pole in the dressed quark propagator. Formulated in Euclidean space, the dynamics is driven by a pure Lorentz-vector kernel, essentially a dressed gluon propagator.

The CST belongs to the approaches related to the BSE, but is similar in spirit to the DS-BS framework in that it aims to incorporate the dynamical origin of the constituent quark masses by dressing the bare quark propagators with the interquark kernel in a consistent fashion. However, the CST is formulated and solved directly in Minkowski momentum space. This is advantageous over Euclidean formulations (although a number of singularities have to be handled numerically) because no analytic continuations are needed to calculate, e.g., form factors \cite{pionff:2014,pionff:2015}, even in the timelike region. The chosen interaction kernel is a manifestly covariant generalization of the Cornell potential, and the full Dirac structure of the quarks is taken into account. 

\begin{figure}[tb]
\captionsetup[subfloat]{singlelinecheck=off}
\centering
  \subfloat[]{%
    \includegraphics[width=.3\textwidth]{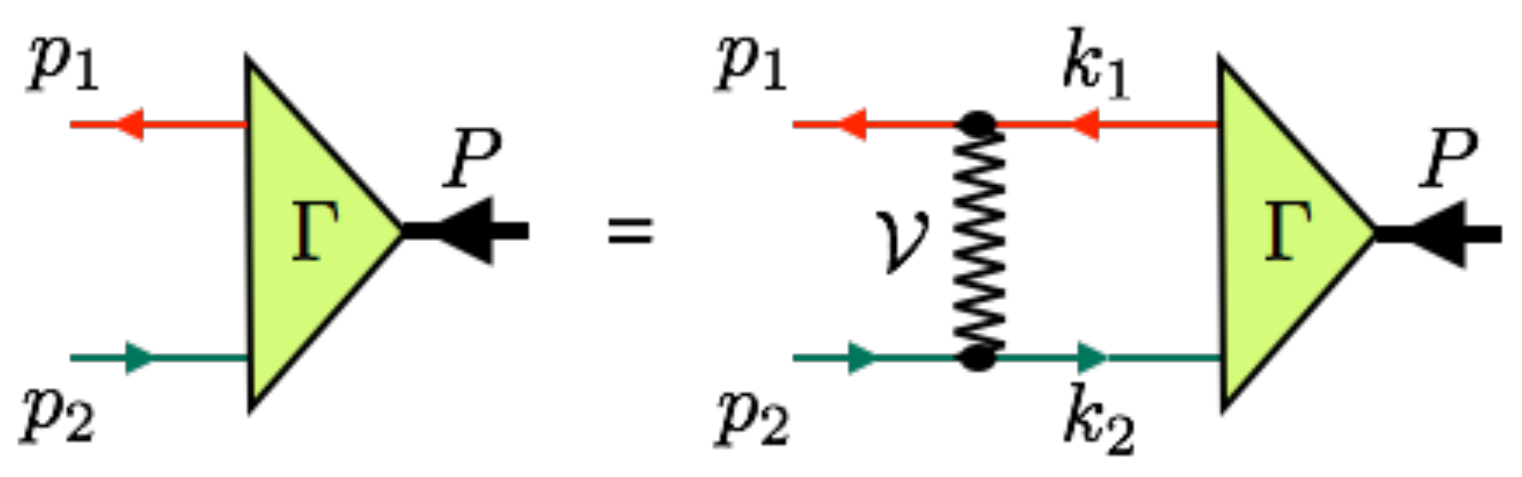}}\\
  \subfloat[]{%
    \includegraphics[width=.48\textwidth]{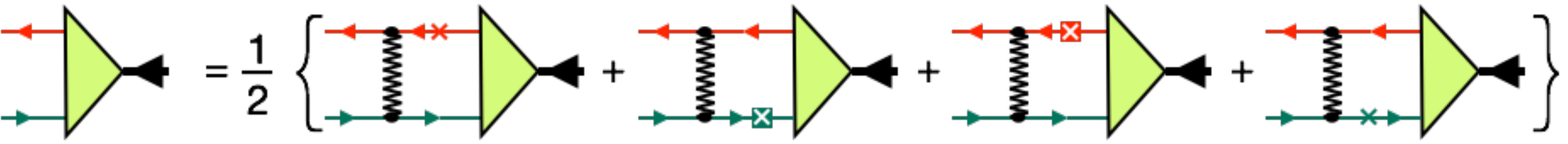}}\\
  \subfloat[]{%
    \includegraphics[width=.48\textwidth]{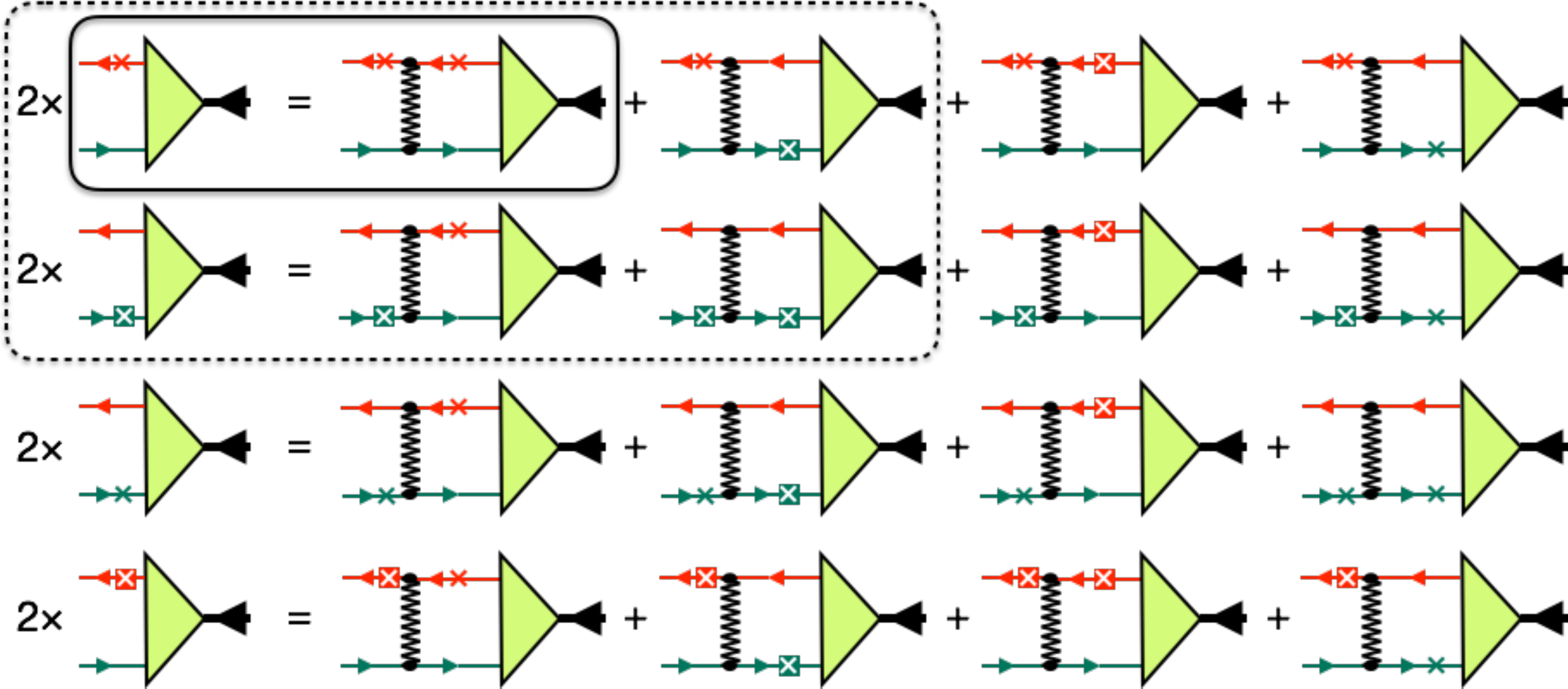}}
  \caption{Graphic representations of (a) the BSE for the $q\bar{q}$ bound state vertex function $\Gamma$, where ${\cal V}$ represents the kernel of two-body irreducible Feynman diagrams; (b) the BS vertex function approximated as a sum of CST vertex functions (crosses on quark lines indicate that a positive-energy pole of the propagator is calculated, light crosses in a dark square refer to a negative-energy pole);  (c) the complete CST equation. The solid rectangle indicates the one-channel equation used in this work, the dashed rectangle a two-channel extension with charge-conjugation symmetry.}\label{fig:1}
\end{figure}

The Covariant Spectator Equation (CSE) is obtained from the BSE [Fig.~\ref{fig:1}(a)] by carrying out the energy loop integration such that only quark-propagator pole contributions are kept [Figs.~\ref{fig:1}(b) and \ref{fig:1}(c)]. This prescription is motivated by partial cancellations between higher-order ladder and crossed-ladder kernels, implying that a CST ladder series effectively contains crossed-ladder contributions which are necessary for the two-body equation to reach the correct one-body limit \cite{Gross:1982}.

%
%

In this work we are focussing on systems where one quark is typically much heavier than the other, so we are close to the one-body limit. The BS ladder approximation does not possess this limit, and it would not be a good choice to describe these mesons. On the other hand, heavy-light systems are ideal to apply a simplified version of the CSE, the so-called one-channel spectator equation (1CSE): the positive-energy pole of the heavier quark dominates, such that the other three CST vertex functions can be neglected. The 1CSE is shown in Fig.~\ref{fig:1}(c), inside the solid rectangle.

This equation retains most important properties of the complete CSE, namely manifest covariance, cluster separability, and  the correct one-body limit. It is also a good approximation for equal-mass particles, as long as the bound-state mass is not too small (this excludes the pion from its range of applicability). In fact, in a properly symmetrized form to account for the Pauli principle, it has been applied very successfully to the description of the two- and three-nucleon systems \cite{Gro08,Gro10,Sta97}. 

A property the 1CSE does not maintain is charge-conjugation symmetry. Therefore, heavy quarkonium states calculated with the 1CSE have no definite C-parity. In principle, this problem is easily remedied by using instead the two-channel extension inside the dashed rectangle of Fig.~\ref{fig:1}(c). However, we decided that the considerable increase in computational effort would not be justified for the purpose of this work: of the quarkonia with $J^P=0^\pm$ and $1^\pm$, only the axial-vector mesons ($J^P=1^+$) come in both C-parities, and these pairs are separated by only a few MeV (5 to 6 MeV in bottomonium, 14 MeV in charmonium). Thus, as long as we do not seek an accuracy better than about 10-20 MeV, the use of the 1CSE also for heavy quarkonia is perfectly justified. Consistent with this level of accuracy, we also set $m_u=m_d$ throughout this work.

We use a kernel of the general form
\begin{multline}
{\cal V}= \left[ (1-y) \left({\bf 1}_1\otimes {\bf 1}_2 + \gamma^5_1 \otimes \gamma^5_2 \right) - y\, \gamma^\mu_1 \otimes \gamma_{\mu 2} \right]V_\mathrm{L}  
\\
-\gamma^\mu_1 \otimes \gamma_{\mu 2} \left[ V_\mathrm{OGE}+V_\mathrm{C} \right] 
\equiv
\sum_K V_K  \Theta_{1}^{K(\mu)} \otimes \Theta^K_{2(\mu)}\, ,
\label{eq:kernel}
\end{multline}
where $V_\mathrm{L}$, $V_\mathrm{OGE}$, and $V_\mathrm{C}$ are relativistic generalizations of a linear confining potential, a short-range one-gluon-exchange (in Feynman gauge in this work), and a constant interaction, respectively.  The confining interaction has a mixed Lorentz structure, namely equally weighted scalar and pseudoscalar structures, and a vector structure. The parameter $y$ dials continuously between the two extremes, $y=1$ being pure vector coupling, and $y=0$ pure scalar+pseudoscalar coupling. The OGE and constant potentials are Lorentz-vector interactions. The signs are chosen such that---for any value of $y$---in the static nonrelativistic limit always the same Cornell-type potential $V(r)=\sigma r -\alpha_s/r -C$ is recovered.

The reason for the presence of a pseudoscalar component is chiral symmetry. Although in general scalar interactions break chiral symmetry, it was shown in ~\cite{CSTpi-pi} that the CSE with our relativistic linear confining kernel satisfies the axial-vector Ward-Takahashi identity when it is accompanied by an equal-weight pseudoscalar interaction. It has also been shown \cite{Gross:1991te,CST:2014} that, in the chiral limit of vanishing bare quark mass, a massless pion solution of the CSE emerges, while a finite dressed quark mass is dynamically generated by the interaction kernel through a NJL-type mechanism. 

For simplicity, and to establish a reference calculation, we use fixed instead of dynamical, momentum-dependent consituent quark masses in this work. For the same reason, we postpone the inclusion of a running coupling in $V_\mathrm{OGE}$ and use a fixed value of $\alpha_s$ instead.

The 1CSE with quark 1 on its positive-energy mass shell can be written in manifestly covariant form
\begin{multline}
\Gamma(\hat{p}_1,p_2) = - \int \frac{d^3k}{(2\pi)^3} \frac{m_1}{E_{1k}} \sum_K V_K(\hat{p}_1,\hat{k}_1) \Theta_{1}^{K(\mu)}  
\\
\times 
\frac{m_1+\hat{\slashed{k}}_1}{2m_1} \Gamma(\hat{k}_1,k_2)
\frac{m_2+\slashed{k}_2}{m_2^2-k_2^2-i\epsilon}\Theta^K_{2(\mu)} \, ,
\label{eq:1CSE}
\end{multline}
where $\Theta_i^{K(\mu)}={\bf 1}_i, \gamma^5_i,$ or $\gamma_i^\mu$,  $V_K(\hat{p}_1,\hat{k}_1)$ describes the momentum dependence of the kernel $K$, $m_i$ is the mass of quark $i$, and $E_{ik}\equiv \sqrt{m_i^2+{\bf k}^2}$. A ``$\hat{\phantom{p}}$'' over a momentum indicates that it is on its positive-energy mass shell.

The kernel functions $V_K(\hat{p}_1,\hat{k}_1)$ in (\ref{eq:1CSE}) are
\begin{multline}
 V_\mathrm{L}(\hat{p}_1,\hat{k}_1)  = -8\sigma \pi\left[\frac{1}{(\hat{p}_1-\hat{k}_1)^4} -\frac{E_{p_1}}{m_1}(2\pi)^3 \delta^3 (\mathbf{p}_1-\mathbf{k}_1)
 \right.
\\
\left.  \times  \int \frac{d^3 k'_1}{(2\pi)^3}\frac{m_1}{E_{k'_1}}
\frac{1}{(\hat{p}_1-\hat{k}'_1)^4}\right] \, ,
\end{multline}
 
\begin{equation}
V_\mathrm{OGE}(\hat{p}_1,\hat{k}_1)  = - \frac{4 \pi \alpha_s}{(\hat{p}_1-\hat{k}_1)^2} \, ,
\end{equation}

\begin{equation}
V_\mathrm{C}(\hat{p}_1,\hat{k}_1) = (2\pi)^3\frac{E_{k_1}}{m_1} C \delta^3 (\mathbf{p}_1-\mathbf{k}_1) \, .
\label{eq:V}
\end{equation}

Instead of solving (\ref{eq:1CSE}) directly for the vertex functions, we introduce relativistic ``wave functions'' with definite orbital angular momentum, defined as rather complicated combinations of spinor matrix elements of the vertex function multiplied by the off-shell quark propagator \cite{GMilana:1994}. They enable us to determine the spectroscopic identity of our solutions, which is indispensable when comparing to the measured states. In the nonrelativistic limit, they reduce to the familiar Schr\"odinger wave functions. However, our relativistic wave functions contain components not present in nonrelativistic solutions. For example, the $S$-waves of our pseudoscalar states couple to small $P$-waves (with opposite intrinsic parity) that vanish in the nonrelativistic limit, whereas, for vector mesons, coupled $S$- and $D$-waves are accompanied by relativistic singlet and triplet $P$-waves.

The 1CSE for the relativistic wave functions can be written as a generalized linear eigenvalue problem for the total bound-state mass. We solve this system by expanding the wave functions in a basis of $B$-splines, as described in \cite{GMilana:1994,Uzzo}. Special attention is needed to treat the singularities in the kernel at $(\hat{p}_1-\hat{k}_1)^2=0$. We apply techniques similar to the ones described in \cite{Leitao:2014} to obtain stable results. A standard Pauli-Villars regularization is applied to divergent loop integrations, at the expense of a momentum cut-off parameter $\Lambda$. Our results are quite insensitive to the exact value of $\Lambda$, and we simply fix it at twice the heavier quark mass. 

We calculated the pseudoscalar, scalar, vector, and axial-vector meson states that contain at least one heavy (bottom or charm) quark, and whose mass falls below the corresponding open-flavor threshold. As exceptions, a few states 
slightly above threshold but with very small widths are considered as well.

The model parameters are the four constituent quark masses $m_u=m_d$, $m_s$, $m_c$, and $m_b$,  the two coupling strengths $\sigma$ and $\alpha_s$, the constant $C$, and the mixing parameter $y$. Early results clearly favored pure scalar+pseudoscalar confinement, so throughout this work we set $y=0$.

\begin{figure*}[tbp]
\includegraphics[width=0.98\textwidth]{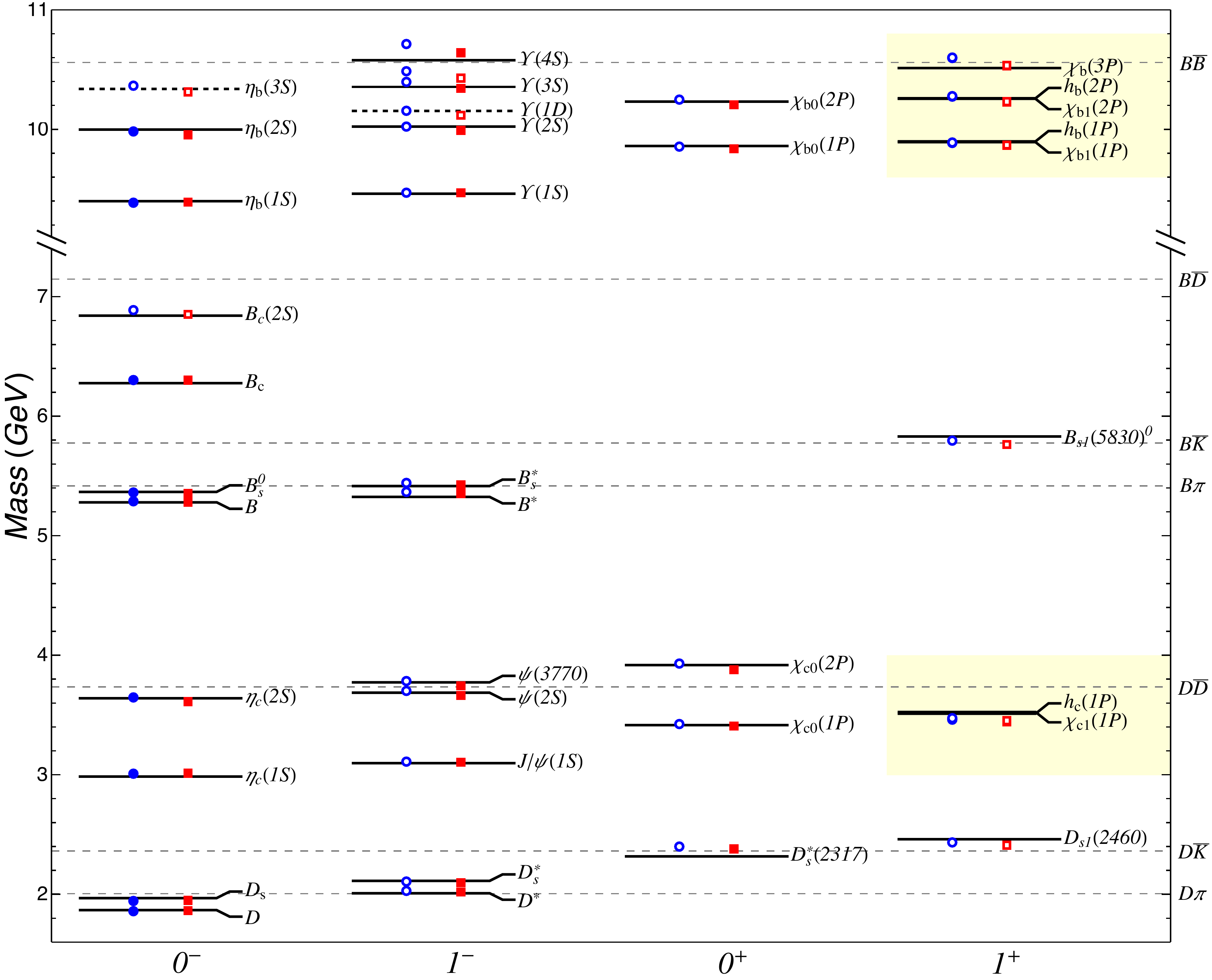}
\caption{Masses of heavy-light and heavy mesons with $J^P=0^\pm$ and $1^\pm$. Circles are 1CSE results with model P1, squares of model PSV1. Solid symbols represent states used in the model fits, open symbols are  predictions. Solid horizontal lines are the measured meson masses \cite{PDG}. The two dashed levels are estimates taken from Ref.~\cite{Godfrey+Moats}. There is weak evidence (at $1.8\sigma$) that the $\Upsilon(1D)$ has been seen \cite{Bonvicini:2004fr,Amo-Sanchez:2010jk}. Both models predict a so far unobserved $\Upsilon(2D)$ between $\Upsilon(3S)$ and $\Upsilon(4S)$.  Dashed horizontal lines across the figure indicate open flavor thresholds. The axial-vector quarkonia are shaded because the 1CSE does not define a specific C-parity for these states.}
\label{fig:spectrum}
\end{figure*}

Figure~\ref{fig:spectrum} shows the results of two different model calculations with the 1CSE in comparison to the observed meson masses. Model P1 was fitted to 9 pseudoscalar states only, whereas model PSV1 was fitted to the masses of 25 pseudoscalar, scalar, and vector mesons. A solid circle (square) in Fig.~\ref{fig:spectrum} indicates a mass calculated with model P1 (PSV1) that was used in the fit to the measured masses (solid lines), whereas the open symbols show predictions of the respective models. The parameters of the models are listed in Tab.~\ref{tab:parameters}. Fitting the quark masses is much more time-consuming than fitting the other parameters. Therefore, we first determined them in preliminary calculations and then held them fixed in the final fits of $\sigma$, $\alpha_s$ and $C$. This procedure is certainly good enough for the purpose of this work.

\begin{table}[tb]
\caption{Kernel parameters of models P1 and PSV1. Both models use the quark masses $m_b=4.892$ GeV, $m_c=1.600$ GeV, $m_s=0.448$ GeV, and $m_u=m_d=0.346$ GeV.}
\begin{center}
\begin{tabular}{l|ccc}
Model & $\sigma$ [GeV$^2$] & $\alpha_s$ & $C$ [GeV] \\
\hline
P1      & 0.2493 & 0.3643 & 0.3491 \\
PSV1 & 0.2247 & 0.3614 & 0.3377 \\
\end{tabular}
\end{center}
\label{tab:parameters}
\end{table}

Figure~\ref{fig:spectrum} clearly shows that both models give results in very good agreement with the experimental meson spectrum. 
It is remarkable that a simple unified model with global parameters $\sigma$, $\alpha_s$, and $C$ can describe heavy-light and heavy mesons over such a large range of masses (calculations in the literature often vary model parameters from sector to sector).

The rms differences to the measured masses are $0.036$ GeV for P1 and $0.030$ GeV for PSV1, which is comparable to the typical rms deviations reported in \cite{Spence:1993bh}. 
The axial-vector states in the shaded area of Fig.~\ref{fig:spectrum} are not considered in these numbers because, as explained above, the predicted states cannot be uniquely identified with the observed C-parity eigenstates. Nevertheless, it is very pleasing that both models predict two tightly-spaced states (the symbols overlap in the figure) in close proximity to the experimental C-parity pairs, both in charmonium ($\chi_{c1}(1P)$ and $h_c(1P)$)  and in bottomonium ($\chi_{b1}(1P)$ and $h_b(1P)$, as well as $\chi_{b1}(2P)$ and $h_b(2P)$).

The fact that the fit exclusively to pseudoscalar mesons of model P1 yields almost the same result as the more general fit of model PSV1 allows us to draw a more fundamental conclusion: the \emph{covariant form} of the kernel (\ref{eq:kernel}) is responsible for a \emph{correct prediction} of the spin-dependent interactions. 

It is of course very well known that the Lorentz structure of a kernel determines the spin-dependent interactions (see, e.g., \cite{Lucha:1991nr}), and it is certainly one of the many attractive features of a covariant formalism that they are not treated perturbatively but on an equal footing with the spin-independent interactions. But in a general fit to all types of mesons one cannot really \emph{test} the predictive power of the covariant kernels in this regard because all interactions are fitted simultaneously. 

However, pseudoscalar states are nearly pure $S$-waves (the tiny relativistic $P$-wave admixture has almost no effect), such that spin-orbit and tensor forces are exactly zero, whereas the spin-spin interaction is probed only in singlet states and cannot be separated from the central force. A fit in which the spin-dependent interactions are completely unconstraint is therefore the ideal case to test their prediction. 

To summarize, from our results one can conclude not only that our covariant kernel is a very efficient way to derive the spectrum and wave functions of heavy-light and heavy mesons, but also that it correctly predicts the spin-dependent interactions solely based on their relation to the spin-independent interactions as dictated by covariance.

\section*{Acknowledgements}
We thank F.\ Gross and J.\ E.\ Ribeiro for helpful discussions, and the Jefferson Lab Theory Group for their hospitality during recent visits when part of this work was performed.
This work was supported by the Portuguese \emph{Funda\c c\~ao para a Ci\^encia e a Tecnologia} (FCT) under contracts SFRH/BD/92637/2013, SFRH/BPD/\-100578/\-2014, and UID/FIS/0777/2013.


\end{document}